\documentclass[prl,twocolumn,aps,amsmath,amssymb,superscriptaddress]{revtex4}
\usepackage{epsfig}
\usepackage{subfigure}
\usepackage{graphicx}
\usepackage{bbm}
\begin{document}

\title{Stability Properties of Nonhyperbolic Chaotic Attractors under Noise} 
\def\active{0} 

\author{Suso Kraut}
\affiliation{Instituto de F\'isica, Universidade de S\~ao Paulo, Caixa Postal
66318, 05315-970 S\~ao Paulo, Brazil} 

\author{Celso Grebogi}
\affiliation{Instituto de F\'isica, Universidade de S\~ao Paulo, Caixa Postal
66318, 05315-970 S\~ao Paulo, Brazil}  
\affiliation{Max-Planck-Institut f\"ur Physik komplexer Systeme, N\"othnitzer
Strasse 38, 01187 Dresden, Germany}

\begin{abstract}
We study local and global stability of nonhyperbolic chaotic attractors
contaminated by noise. The former is given by the maximum distance of a 
noisy trajectory from the noisefree attractor, while the latter is provided 
by the minimal escape energy necessary to leave the basin of attraction,
calculated with the Hamiltonian theory of large fluctuations. We establish the
important and counterintuitive result that both concepts may be opposed to
each other. Even when one attractor is globally more stable than another one,
it can be locally less stable. Our results are exemplified with the Holmes
map, for two different sets of parameter, and with a juxtaposition of the
Holmes and the Ikeda maps. Finally, the experimental relevance of these
findings is pointed out.
\end{abstract}

\maketitle
\begin{figure}[htb]
\begin{center}
\epsfig{file=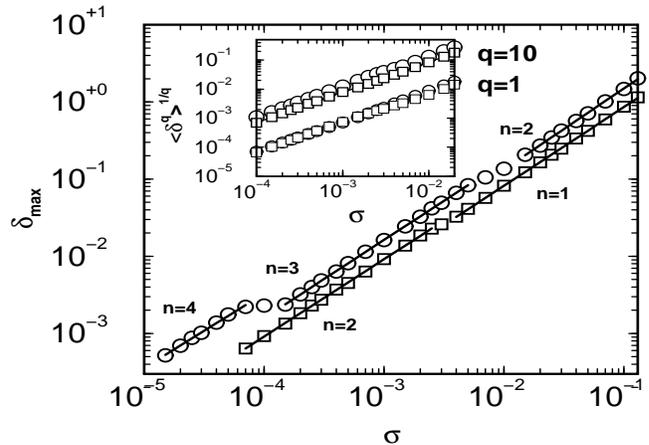,width=8cm,height=6cm}
\caption{Maximum distance $\delta_{max}$ versus noise intensity $\sigma$ for 
the Holmes map, with the parameters (i) $a=0.047$, $b=2.4$ and $c=0.155$ 
(circles) and (ii) $a=0.01$, $b=2.8$ and $c=0.8$ (squares). The regions of
linear growth are fitted by straight lines and marked with the number $n$
of the corresponding image of the PHT. For each noise strength $5 \times 10^9$
iterations have been used. The inset shows the average $\langle \delta ^q
\rangle ^{1/q}$, $q=1, 10$, with Gaussian noise for set (i) (circles) and set
(ii) (squares).  $10^9$ iterations for each noise level have been averaged
over.}     
\label{Holmes}
\end{center}
\end{figure}

Noise plays an important role in nonlinear systems. Specifically, the
fundamental question of the effect of noise on the stability of a chaotic
attractor can be viewed under two  different angles. The first aspect is to
consider the escape from an attractor through random fluctuations. This is
termed {\it global stability}. Relevant examples range from switching in
lasers \cite{Hales:2000}, Penning traps \cite{Lapidus:1999}, over chemical
reactions \cite{Gillespie:1977} to electronic circuits \cite{Luchinsky:1997}. 
Since the seminal work of Kramers \cite{Kramers:1940}, this problem has been
treated for a broad range of settings \cite{Hanggi:1990}. For nonequilibrium
systems, a WKB-like extension of Kramers' equilibrium  theory has been devised
\cite{Onsager:1953,Freidlin:1984}. This so-called  Hamiltonian theory of large
fluctuations uses an approach similar to path integrals, thus obtaining the
{\it most probable exit path} (MPEP). The MPEP, with an exponentially favoured
probability of occurrence, yields in turn the optimal fluctuations and the
minimal escape energy as well.   
\\
\indent
This theory has been employed for the calculation of the escape from a 
periodic state \cite{Kautz:1987,Beale:1989,Grassberger:1989,Dykman:1990,
Silchenko:2003}. Recently, it has also become possible to treat the escape
from a {\it nonhyperbolic chaotic attractor} (NCA) \cite{Kraut:2004}, whose
stable and unstable manifolds exhibit tangencies. It was demonstrated 
that the MPEP is uniquely determined by the {\it primary homoclinic tangency}
(PHT) closest to the basin boundary. A tangency is homoclinic if both manifolds
belong to the same periodic orbit and primary, if a perturbation is
amplified under forward and backward iteration of the dynamics. Since, in 
practice, virtually all chaotic attractors appear to be nonhyperbolic, it can
be considered as the general case.  
\\
\indent
The second aspect of noise effects on NCAs is {\it local stability}, which 
is a measure of the maximum distance of a noisy trajectory from the noisefree
attractor. Here, the trajectory is {\it always} close to the attractor, without
leaving its basin of attraction. The concept of local stability of a NCA 
against noise is of fundamental importance and has bearings, e.g., on noise
reduction \cite{Hammel:1990}, reconstruction of dynamical quantities
\cite{Kostelich:1993}, parameter estimation \cite{McSharry:1999}, noise level
evaluation \cite{Heald:2000}, and communication with chaos \cite{Bollt:1997}. 
When applying noise bounded by $\sigma$, for hyperbolic attractors the 
maximum distance scales as $\delta_{max} \sim \sigma$ \cite{Ott:1985}. For 
NCAs, however, it was shown that there is a much larger $\delta_{max}$ as 
compared to the hyperbolic case \cite{Jaeger:1997}, caused by attractor
elongating deformations along the PHT and their images (see
\cite{Schroer:1998,Kantz:2002}, as well). This was also confirmed
experimentally \cite{Diestelhorst:1999}.     
\\
\indent
In this Letter we contrast these two measures of stability. While it is usually
assumed that they behave in a similar fashion, we point here out, however, 
the counterintuitive effect that a nonhyberpolic chaotic attractor can be, in 
the above defined sense, globally more stable than another one, yet locally 
less stable. This is all the more surprising as both stability properties are
intimately related to the primary homoclinic tangency. This phenomenon can be 
understood, though, by taking into account that for global stability the 
preimages are most relevant, constituting the proper and unique initial
conditions for the most probable exit path \cite{Kraut:2004}. On the other 
hand, for local stability only the images govern the process 
\cite{Jaeger:1997}, as their local expansion rates, given by Eq. 
(\ref{growthrate}) below, contribute to a divergence from the attractor. 
Consequently, for local stability only linear properties of the system are 
relevant, whereas global stability can only be fully described by the complete
set of variational equations, which are nonlinear. 
\\
\indent
We illustrate these findings first with the Holmes map \cite{Holmes:1979} with
two different sets of parameters. Thereafter, we demonstrate this phenomenon
by comparing the Holmes and the Ikeda map \cite{Ikeda:1979}. Since the
Hamiltonian theory of large fluctuations is only valid for Gaussian noise and
the maximum distance is only well defined for bounded noise, we calculate
for local stability also the averaged Gaussian distance, including higher
moments. This removes any particularity of comparing different noise
distributions. The outcome of the calculation corroborates our main claim,
too.     
\\
\indent
As a fundamental dynamical example we consider the Holmes map 
\cite{Holmes:1979}
\begin{eqnarray}
\begin{split}
x_{n+1} & = \,y_n + \xi_{x} & \\
y_{n+1} & = \,a\,x_n + b\,y_n  -  c\,y_n ^3 + \xi_{y}, & 
\end{split}
\end{eqnarray}
with the white noise terms  $\xi_{x}, \xi_{y}$ uniformly distributed in 
the disk $\sigma$: $\xi_{x}^2 + \xi_{y}^2 \le \sigma$.
We choose the first set of parameters to be (i) $a=0.047$, $b=2.4$ and 
$c=0.155$. That gives two attractors, symmetrical with respect to the
origin; we focus only on one of these. When increasing $b$, these attractors
merge in a crisis. Our second set of parameters is then in the region, where
only one large symmetric chaotic attractor exists, (ii) $a=0.01$, $b=2.8$ 
and $c=0.8$. Both NCAs are normalized in a twofold way. First, the extensions 
in phase space $E = \sqrt{(x_{max}-x_{min}) (y_{max}-y_{min})}$ are demanded
to be the same, because then the percentage of noise on each attractor is
identical. This is a common measure of the relative noise intensity, in turn
adjusting the local properties. Second, the threshold of escape from the NCAs
with bounded noise is also required  to be equal. This guarantees the same
scaling region for the maximum distance and calibrates the global
properties. For the chosen parameters, the two measures yield $E \approx 3.1$
and $\sigma_{escape} \approx 0.13$ \cite{footnote1}. With these two conditions
met, the comparison is as general and unambiguous as possible.  
\\
\indent
Let $\delta_{k}$ be at each step of iteration the minimum distance of the 
noisy trajectory from the noiseless attractor. The maximum distance
$\delta_{max}$ of the whole trajectory is then defined as the maximum over all
the minimum distances: $\delta_{max} = \underset{k}{\max} \,(\delta_k)$. 
For the numerical computation, we partition the attractor with
a grid of box edge length $l$, where $l$ depends on the noise strength and 
the desired resolution ($0.0005 \le l \le 0.01$). We store only a limited
number of points of the noiseless attractor per box of the grid
(ca. 100). Each point of the noisy trajectory is then compared solely to
attractor points of the box it falls in and the neighboring ones. If they
are empty, the number of neighbors is increased until a point of the attractor
has been encountered. This provides, for each trajectory point, the minimum
distance from the attractor $\delta_{k}$, and the largest of these is 
$\delta_{max}$. With this method we get a much better accuracy and a larger
scaling region than in \cite{Schroer:1998,Kantz:2002}, while simultaneously
saving storage and computation time.    
\\
\indent
The result of the calculation for the two NCAs is shown in Fig. \ref{Holmes}.
The scaling is limited for small noise by our computational resolution and 
for large noise by the trajectory escaping from the attractor. It is apparent
from the graph that, for all noise intensities, set (i) (circles) exhibits a
larger $\delta_{max}$ than set (ii) (squares), indicating that the attractor
(ii) is locally more stable than (i).
\\
\indent
The two curves in the log-log scale of Fig. \ref{Holmes} are straight lines,
interrupted by bends. Between two bends, they have an identical slope 1 
(i.e. $\delta_{max} = P_n \,\sigma$). The factor of proportionality $P_n$ 
varies with $n$, causing a different offset. The $\delta_{k}$ achieve their
maxima at the PHT and images thereof (see Fig. 5 of \cite{Jaeger:1997} for a
very instructive illustration). At the $n-th$ image, a perturbation at the PHT
grows like \cite{Jaeger:1997} 
\begin{equation}
P_{n} = \sum_{i=0}^{n-1} \left| \prod_{j=1}^{n-i} \{ Df[f^{i-1+j}({\bf x})]\} 
{\bf e}_{n-i}^{me} [ f^{j}({\bf x}) ] \right| + 1,
\label{growthrate}
\end{equation}
where $f({\bf x})$ is the dynamics of the system at ${\bf x}$, $Df({\bf x})$ 
the Jacobian, and ${\bf e}_{n-i}^{me}[f^{j}({\bf x})]$ the most expanding unit 
vector at $f^{j}({\bf x})$ under the application of $Df^{n-i}({\bf x})$.
This factor sums up all the maximal stretching factors of the PHT and its
images up to the $n-th$ one. Typically, $P_{n} > P_{m}$ for $n > m$, implying 
that the distance from the attractor increases with the number of images. 
However, because higher images are folded back on the attractor, other
parts of the attractor instead of an iteration of the PHT come closer to the
noisy trajectory as the noise level is incremented. This results in a
saturation of the maximum distance, which produces a bend. In turn, when the
noise strength is further augmented, $\delta_{max}$ switches to the next lower
image of the PHT, again initiating a regime of linear growth, and so on. 
\\
\indent
In Fig. \ref{Holmes}, the sequences $4 \rightarrow 3 \rightarrow 2$ for set 
(i) and $2 \rightarrow 1$ for set (ii) can be seen. For set (ii), the offsets
from the numerics of  Fig. \ref{Holmes} for $n = 1, 2 $ are $8.3, 9.5$ (solid
lines), while Eq. (\ref{growthrate}) yields $P_n = 8.2, 9.5$, a very good
agreement. Set (i) for $n = 2 , 3 , 4$ gives $15, 17, 34$ from Fig. 
\ref{Holmes} (solid lines), whereas Eq. (\ref{growthrate}) results in $P_n =
13.4, 16, 34$, also a reasonably good agreement. The values for lower images
of the PHT (i. e. higher noise) fit slightly worse. However, the matching can
be improved by using the full dynamics instead of the linearized Eq. 
(\ref{growthrate}), since nonlinear effects play an increasing role for
larger noise levels. By doing this, one gets  $ 14.5 , 16.5, 34$, again in 
good accordance. 
\\
\indent
To provide a better basis for the comparison with global stability, we 
calculate the averaged moments of the distance $\langle \delta ^q \rangle 
^{1/q}$ = $ (\frac{1}{N}\sum_{k=1}^{N}(\delta_k)^q)^{1/q}$ using Gaussian
white noise, with $\langle\xi_i\rangle=0$ and $\langle\xi_i,\xi_j\rangle=
\sigma^2\, \delta_{ij}$. This is shown in  Fig. \ref{Holmes}, inset, for $q =
1, 10$. The corresponding moments for set (i) are for all $q$ above the ones
of set (ii), more distinctive for higher q. The same applies for $\langle
\delta ^q \rangle^{1/q}$ with bounded noise (not shown). Here, in the limit $q
\rightarrow \infty$ the maximum distance is recovered $\langle \delta ^q
\rangle ^{1/q} \rightarrow \delta_{max}$.
\\
\indent
Global stability is evaluated with the Hamiltonian theory of large
fluctuations, solving a variational equation for the MPEP  
\cite{Grassberger:1989,Dykman:1990,Silchenko:2003}, which provides the
action $S = \frac{1}{2} \sum_{n=1}^{N}{\bf \lambda_n^T} \,{\bf \lambda_n}$,
with ${\bf \lambda_n}$ the optimal fluctuations. The mean first exit time is 
then given by $\langle\tau \rangle \sim \exp \left[ \frac{S}{\sigma^2}\right]$.
The MPEP starts at the preimages of the PHT, leaves the attractor close to the
PHT and moves along their images towards the saddle point on the basin 
boundary \cite{Kraut:2004}. Employing this scheme, one obtains for set (i) 
$S \approx 0.015$ and for set (ii) $S \approx 0.01$, meaning now that set (i)
is globally more stable than set (ii). We stress that this leads, e.g. for a 
noise value of $\sigma ^2 = 0.001$, to an amplification of $\langle \tau 
\rangle$ by a factor of $\exp \left[\frac{0.005} {\sigma^2}\right]\approx148$,
it is therefore no small effect.  
\\
\indent
These opposing stability properties establish our main result. The Holmes map 
(as a typical example of a NCA) is with set (i) of parameters locally less
stable than with set (ii), i.e., the maximum distance $\delta_{max}$ is
larger, but globally more stable, i.e., the escape energy and consequently the
mean first exit time are larger. 
\begin{figure}[htb]
\begin{center}
\epsfig{file=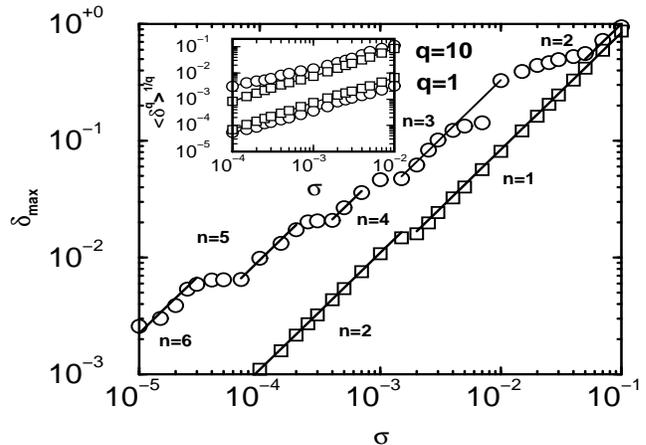,width=8cm,height=6cm}
\caption{Maximum distance $\delta_{max}$ versus noise level $\sigma$ for 
the Ikeda and the Holmes maps, with the parameters (Ikeda) $a=0.9, b=0.9,
\kappa = 0.3$, and $\eta =6.0$ (circles) and set (iii) $a=0.01$, $b=2.78$ and 
$c=1.56$ (squares). For each noise strength $5 \times 10^9$ iterations have
been used. The inset shows the average $\langle \delta ^q \rangle ^{1/q}$, 
$q=1, 10$, with Gaussian noise for Ikeda (circles) and set (iii) (squares). 
$10^9$ iterations for each noise level have been averaged over.}   
\label{Ikeda}
\end{center}
\end{figure}
\\
\indent
Next we demonstrate that this phenomenon can be much more pronounced when 
comparing two NCAs originating from different dynamical systems. For that
purpose, we introduce the Ikeda map \cite{Ikeda:1979}
\begin{equation}
{\bf z_{n+1}} = a + b \,{\bf z_n} \exp \left[ i \kappa - \frac{i \eta}{1 + 
|{\bf z_n}|^2} \right] + {\bf \xi_n},
\label{Ikeda_map}
\end{equation}
where ${\bf z_n} = x_n + i y_n$. We fix the parameters at $a = 0.9, b = 0.9,
\kappa = 0.3$ and $\eta =6.0$, which results in a NCA. We compare this NCA
with the one obtained for the Holmes map with the parameter set (iii) 
$a=0.01$, $b=2.78$ and $c=1.56$. Again both attractors are normalized in the
two ways explained above, with $E \approx 2.1$ and $\sigma_{escape} \approx
0.1$ \cite{footnote2}. The maximum distance is depicted in Fig. \ref{Ikeda}. 
The features are more striking than in the previous example, $\delta_{max}$ 
differs, for instance, for $\sigma = 10^{-4}$, by one order of magnitude. 
Furthermore, the scenario of jumping from one image of the PHT to the next 
one happens for the Ikeda map more frequently. For the lowest noise level
considered, the maximum distance occurs at the $6-th$ image of the PHT. 
\\
\indent
For set (iii) of the Holmes map, the numerical offsets of Fig. \ref{Ikeda}
(solid lines) come about as $8.2, 11$ for the images $n = 1, 2 $ of the
PHT. Equation (\ref{growthrate}) results in $P_n = 8, 11.25$, agreeing 
extremely well. The Ikeda map gives for $n = 2, 3, 4, 5, 6$ the numerical 
values $10, 33, 53, 95, 225$, respectively, while evaluated with
Eq. (\ref{growthrate}) produces $P_n =  8, 22, 50, 96, 247$. Taken into
account that on the one hand for large images of the PHT the maximal distance
is numerically hard to observe, as several subsequent optimal fluctuations are
needed to achieve it, and on the other hand for low images the noise is
already so large as to cause nonlinear effects, the correspondence is
tolerably good. 
\\
\indent
In the inset of Fig. \ref{Ikeda}, the averaged distances with Gaussian noise
$\langle \delta ^q \rangle ^{1/q}, q=1, 10$ are displayed. For small $q$ the
Holmes map is here above the Ikeda map. This is rooted in the fact that the
unstable manifold of the Ikeda map is more curved at the PHT and their 
images. Hence, the average exhibits less of the maximal possible expansion. 
However, for larger $q$  ($q=10$ in the graph), the average is above for all
noise levels. Again, the same holds for $\langle \delta ^q \rangle ^{1/q}$ and
bounded noise (not shown). 
\\
\indent
Global stability analysis, as before, entails for the Ikeda map $S \approx 
0.025$ and for the parameter set (iii) of the Holmes map  $S \approx 0.007$.
Consequently, for the selected parameters, the Ikeda map is globally much more
stable than the Holmes map, while it is locally much less stable, which
is caused by the higher images of the PHT having larger expansion factors
[Eq. (\ref{growthrate})] and at the same time weaker folding back to the 
attractor. Both effects are most pronounced in the relevant low noise limit.
The Ikeda map is globally more stable by a factor of $\exp \left[\frac{0.018} 
{\sigma^2} \right]$. The amplification becomes huge for small noise 
(e.g. $6.5 \times 10^7$ for $ \sigma^2 = 0.001$) and is easily measurable.
This establishes that the phenomenon of opposite stability properties, when
comparing NCAs originating from different dynamical models, can be observed in
an even more striking manner. 
\\
\indent
We have confirmed this counterintuitive phenomenon also when comparing 
the H{\'e}non map with both, the Ikeda and the Holmes maps, corroborating our
findings, which we claim to be a general feature of NCAs.
\\
\indent
In the present work, we were not concerned with the overall scaling of 
$\delta_{max}$, only with the fact that one curve lies above another, thus
implying being locally less stable. In \cite{Schroer:1998}, however, it was
claimed that the scaling is $\delta_{max} \sim \sigma^{\gamma}$, were $\gamma =
1/D_1$, with  $D_1$ the information dimension of the attractor. The 
agreement between this value and our, very accurate, numerics, is not too
good, though \cite{footnote3}. This discrepancy is caused by the fact that in
the derivation of the scaling in \cite{Schroer:1998} not $P_n$ from
Eq. (\ref{growthrate}) was used, but the positive Lyapunov exponent, which
usually has a smaller value. Thus, in general $1/D_1$ can be regarded only as a
lower bound for $\gamma$. The question of an exact scaling exponent will be
treated in \cite{Kraut:2005}.  
\\
\indent
As our findings can have an huge effect on the maximum distance and the
average escape time, they have also relevance for experiments, since one 
cannot simply and straightforwardly conclude the behavior of one of the
stability types by measuring the other. 
\\
\indent
We acknowledge D. G. Luchinsky, S. Beri, A. Pikovsky, M. S. Baptista,
K. M. Zan, R. D. Vilela, A. E. Motter, and H. Kantz for valuable hints and
discussions. This work was supported by the Alexander von Humboldt Stiftung,
CNPq, and FAPESP.


\begin{thebibliography}{10}
\bibitem{footnote1}
For the symmetric case, we only consider half of the attractor for the
calculation of $E$, since it appears to be a more natural way to compare it
with the other, asymmetric one. However, if one wants to take the entire 
attractor, one can choose the parameters $a=0.01$, $b=2.745$ and $c=1.34$. 
This yields again $E \approx 3.1$ and $\sigma_{escape} \approx 0.13 $.
All other results remain qualitatively the same, as well.
\bibitem{footnote3}
We get for sets (i) and (ii) of Fig. \ref{Holmes} $\gamma = 0.89$ and $\gamma
= 0.98$, respectively, while $1/D_1 = 0.89$ and $1/D_1 = 0.87$. For
Fig. \ref{Ikeda} we obtain $\gamma = 0.66$ for Ikeda and  $\gamma = 0.95$ for
set (iii), where $1/D_1 = 0.61$ and $1/D_1 = 0.87$. The largest deviation is
more than 10\%.
\end{thebibliography}
\end{document}